\documentclass[sigconf]{acmart}
\usepackage{enumitem}
\usepackage{pgfplots}
\usepackage{needspace}
\pgfplotsset{compat=1.18}

\setlist[itemize]{leftmargin=*, itemsep=2pt, topsep=2pt, parsep=0pt}
\setlist[enumerate]{leftmargin=*, itemsep=2pt, topsep=2pt, parsep=0pt}
\definecolor{sbblue}{HTML}{4C72B0}
\definecolor{sborange}{HTML}{DD8452}
\AtBeginDocument{%
  }

\setcopyright{none}
\copyrightyear{2026}
\acmYear{2026}

\begin{document}

\title{UniPinRec: Unifying Generative Retrieval and Ranking at Pinterest Scale}


\author{Hanyu Li}
\email{hanyuharryli@pinterest.com}

\authornote{Corresponding authors.}
\affiliation{
  \institution{Pinterest}
  \city{Palo Alto, CA}
  \country{USA}
}

\author{Yi-Ping Hsu}
\email{yhsu@pinterest.com}
\affiliation{
  \institution{Pinterest}
  \city{Palo Alto, CA}
  \country{USA}
}
\author{Aditya Mantha}
\email{amantha@pinterest.com}
\affiliation{
  \institution{Pinterest}
  \city{Palo Alto, CA}
  \country{USA}
}
\author{Prabhat Agarwal}
\email{pagarwal@pinterest.com}
\authornote{Work done at Pinterest.}
\affiliation{
  \institution{Pinterest}
  \city{Palo Alto, CA}
  \country{USA}
}
\author{Laksh Bhasin}
\email{lbhasin@pinterest.com}
\affiliation{
  \institution{Pinterest}
  \city{Palo Alto, CA}
  \country{USA}
}

\author{Jialu Wang}
\email{jialuwang@pinterest.com}
\affiliation{
  \institution{Pinterest}
  \city{Palo Alto, CA}
  \country{USA}
}

\author{Hongtao Lin}
\email{hongtaolin@pinterest.com}
\affiliation{
  \institution{Pinterest}
  \city{Palo Alto, CA}
  \country{USA}
}

\author{Bella Huang}
\email{shuang@pinterest.com} 
\affiliation{
  \institution{Pinterest}
  \city{Palo Alto, CA}
  \country{USA}
}

\author{Yaxin Li}
\email{yaxinli@pinterest.com} 
\affiliation{
  \institution{Pinterest}
  \city{Palo Alto, CA}
  \country{USA}
}

\author{Xinyi Li}
\email{xinyi@pinterest.com}
\affiliation{
  \institution{Pinterest}
  \city{Palo Alto, CA}
  \country{USA}
}

\author{Chuxi Wang}
\email{chuxiwang@pinterest.com}
\affiliation{%
  \institution{Pinterest}
  \city{Palo Alto, CA}
  \country{USA}
}

\author{Kousik Rajesh}
\email{krajesh@pinterest.com}
\affiliation{%
  \institution{Pinterest}
  \city{Palo Alto, CA}
  \country{USA}
}

\author{Hooshmand S. Razaghi}
\email{hshokrirazaghi@pinterest.com}
\affiliation{%
  \institution{Pinterest}
  \city{Palo Alto, CA}
  \country{USA}
}

\author{Shunyao Li}
\email{sli@pinterest.com}
\affiliation{%
  \institution{Pinterest}
  \city{Palo Alto, CA}
  \country{USA}
}

\author{Zongyue Qin}
\email{zqin@pinterest.com}
\authornotemark[2]
\affiliation{%
  \institution{Pinterest}
  \city{Palo Alto, CA}
  \country{USA}
}

\author{Jaewon Yang}
\authornotemark[1]
\email{jaewonyang@pinterest.com}
\affiliation{%
  \institution{Pinterest}
  \city{Palo Alto, CA}
  \country{USA}
}

\author{James Li}
\email{jamesli@pinterest.com}
\affiliation{%
  \institution{Pinterest}
  \city{Palo Alto, CA}
  \country{USA}
}

\author{Dhruvil Deven Badani}
\email{dbadani@pinterest.com}
\affiliation{%
  \institution{Pinterest}
  \city{Palo Alto, CA}
  \country{USA}
}

\author{Jiajing Xu}
\email{jiajing@pinterest.com}
\affiliation{%
  \institution{Pinterest}
  \city{Palo Alto, CA}
  \country{USA}
}

\author{Charles Rosenberg}
\email{crosenberg@pinterest.com}
\affiliation{%
  \institution{Pinterest}
  \city{Palo Alto, CA}
  \country{USA}
}

\renewcommand{\shortauthors}{Li et al.}


\begin{abstract} 
Modern recommendation systems predominantly train retrieval and ranking as separate models despite both increasingly relying on large transformers encoding the same user behavior data, duplicating parameters, compute, and serving cost.
Prior work unifies the model architecture but not the full pipeline: input formats, training procedures, and serving stacks remain fragmented across stages.

We present \textbf{UniPinRec}, which achieves full-stack unification of retrieval and ranking at Pinterest: one input format, one model, one training stage, deployed within existing serving infrastructure.
A shared transformer encodes the user action sequence into candidate-independent representations that branch into retrieval (ANN dot-product) and ranking (cross-attention) via task-specific heads.
Three ideas make this work: (1) Masked Action Modeling (MAM) eliminates interleaving, enabling weight sharing without doubling context length; (2) Blended training examples pair action sequences with feedview impression slates to satisfy both objectives jointly; (3) Cross-stage KV cache sharing reuses user-history computation from retrieval for ranking, reducing total FLOPs versus serving two independent models.

Deployed in the Pinterest core surfaces, UniPinRec delivers approximately +1\% online engagement lift while cutting end-to-end serving latency by 11.1\% and lifting QPS by 63.6\%.
To our knowledge, this is the first full-stack unification of retrieval and ranking, covering inputs, model, training and serving, deployed in a production recommendation system.

\end{abstract}

\ccsdesc[500]{Information systems~Recommender systems}

\keywords{generative recommendation, unified retrieval and ranking, KV-cache reuse}


\maketitle

\section{Introduction}\label{sec:intro}
Modern industrial recommendation systems~\cite{Chen2025PinFM, Agarwal2025PinRec, Transact23} are built as multi-stage funnels: candidate generation retrieves thousands of items from a corpus of millions, ranking scores these candidates with richer features, and blending assembles the final page.
Each stage is trained in isolation with its own architecture, objective, and training data.
This separation carries a fundamental cost: retrieval and ranking both learn representations from the same user behavior data, yet cannot share parameters or transfer learned signals across stages.

\begin{figure*}[!t]
    \centering
    \includegraphics[width=0.95\linewidth, trim=0 0 0 0, clip]{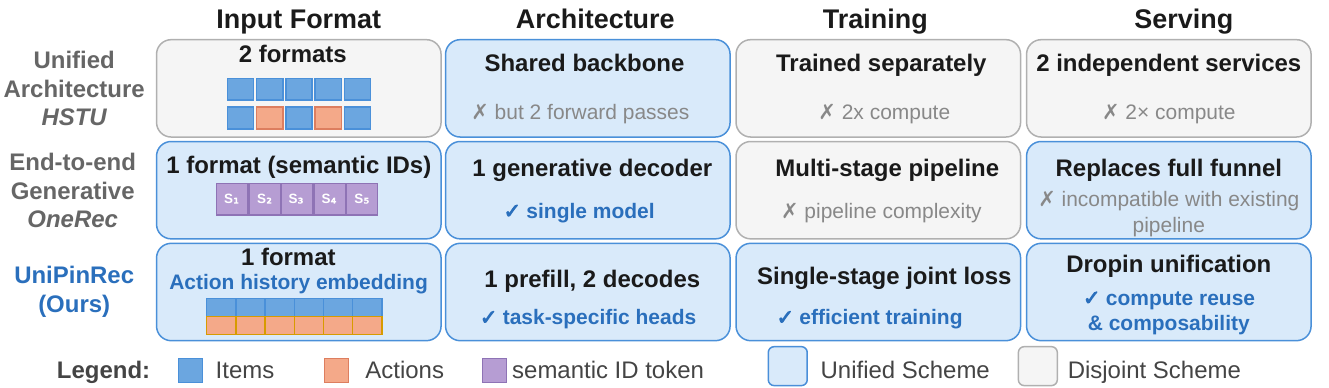}
    \caption{Prior methods unify the architecture but leave input format, training, and serving fragmented. Our approach unifies all four dimensions.}
    \vspace{-4mm}
    \label{fig:overview}
\end{figure*}

The emergence of transformer-based sequence models as the dominant backbone at every stage~\cite{Zhai2024ActionsSpeakLouder, Ye2025FuXiAlpha, Chen2026VISTA, Sun2026OneRanker, Dai2025OnePiece} has made this redundancy more acute.
When both stages are powered by large transformers encoding the same user action history, duplication becomes the primary source of inefficiency: in parameters, training compute, and serving cost.
This creates a concrete opportunity: unify the two stages into a single model that encodes the user once, serves both stages from a common representation, and enables signal sharing where ranking supervision informs retrieval and retrieval's broader corpus exposure regularizes ranking~\cite{Dai2025OnePiece, Sun2026OneRanker}.

We call this goal \emph{full-stack unification} (Figure~\ref{fig:overview}): sharing not just the model architecture, but the entire pipeline, including input format, training, and serving infrastructure.
In this paper, we achieve full-stack unification and deploy it in production at Pinterest.

\paragraph{Why full-stack unification is hard.}
Three challenges have prevented prior work from achieving this.
\textit{(1) Usage divergence:} Although the backbone architecture has converged, the computational regimes remain distinct: retrieval must score millions of candidates under millisecond latency (e.g., via ANN dot-product scoring~\cite{Agarwal2025PinRec}), while ranking scores hundreds and can afford richer per-candidate computation (e.g., cross-attention~\cite{Chen2025PinFM, Zhai2024ActionsSpeakLouder}).
\textit{(2) Training complexity:} Retrieval and ranking use different objectives (sampled softmax over the full corpus vs. binary classification over a small impression set), but prior work on next-action prediction~\cite{Chen2025PinFM, xia2025transactv2} already suggests these tasks could be complementary. Nonetheless, existing unified models resort to multi-stage pipelines (pre-training, fine-tuning, RL alignment) to reconcile them, adding complexity that itself prevents full-stack unification.
\textit{(3) Serving and operational challenges:} A unified model must serve at a lower cost than when separately deployed to justify the change; another desirable property is composability with existing candidate generators (e.g., keywords, trending), and preserve the flexibility to determine the level of unification, and A/B test and roll back each stage independently.
These challenges are largely unexplored; very few systems have deployed a unified retrieval-ranking model in production~\cite{Deng2025OneRec, Sun2026OneRanker}.

\paragraph{Why existing methods fall short.}
Existing approaches fall into two broad directions (Figure~\ref{fig:overview}).
The first proposes \emph{unified architectures} across stages.
HSTU~\cite{Zhai2024ActionsSpeakLouder} uses a shared transformer paradigm but requires different input formats for retrieval (non-interleaved) and ranking (interleaved), so the stages are trained independently and cannot share computation.
OnePiece~\cite{Dai2025OnePiece} adds unified multi-task training but was only deployed as either a retrieval or ranking model, never serving both stages from one unified model.
In short, these methods share the architecture but remain far from full-stack unification.

The second direction proposes \emph{end-to-end generative methods} that bypass the funnel entirely.
OneRec~\cite{Deng2025OneRec} and OneRanker~\cite{Sun2026OneRanker} produce candidates via autoregressive decoding of semantic IDs, achieving unification by construction. 
However, replacing the production funnel entirely makes it harder to compose with existing candidate sources and loses per-stage operational control.

\paragraph{Our approach.}
Our design principle is to compose existing, proven components (ANN indexing, cross-attention ranking, KV caching) rather than introducing new mechanisms, making the unified model a drop-in replacement within existing production infrastructure.
Unification is not merely an elegance goal: a separate ranker would duplicate the expensive user-history encoding that retrieval already performs. By sharing this computation, UniPinRec adds ranking at marginal cost, making it practical to rank within the same latency budget as retrieval alone.

UniPinRec achieves this with three design choices. First, Masked Action Modeling adds ranking supervision to the same non interleaved user sequence used for retrieval. Second, blended training examples and a joint loss train retrieval and ranking in a single stage. Third, cross-stage KV-cache sharing lets ranking reuse the user-history computation from retrieval, preserving modular serving while avoiding duplicate transformer work.

Together, these choices achieve full-stack unification: same input format for both stages, single-stage joint training, and deployment within existing serving infrastructure.

\paragraph{Results.}
The unified model adds ranking capability while maintaining retrieval recall and reducing serving cost, a combination not achievable without model/system co-design.
Offline, UniPinRec lifts ranking Hit@3 by $+14.8\%$ over the production ranker while matching production retrieval recall.

Because UniPinRec couples retrieval and ranking through a shared backbone, the most natural incremental rollout scheme starts from the L0 side: we ship retrieval together with L1 light-weight ranking as a single service (L0+L1), and extend to L2 fine ranking as a follow-up. This keeps the downstream ranker/blender untouched and gives us clean A/B attribution early.

Although the offline ranking lift is attenuated by the downstream ranker/blender, it still delivered consistent engagement wins, $+0.95\%$ on Board More Ideas saves and $+0.91\%$ on Notifications push opens.
On serving, optimizations like cross-stage KV-cache reuse yield a $>3\times$ total ranking forward-pass speedup (Table~\ref{tab:serving_latency}) and in production, -11.1\% e2e latency reduction and +63.6\% QPS lift from naive separate deployment (Table~\ref{tab:serving_qps}).

\section{Related Work}\label{sec:related}
\paragraph{Transformer-based recommendation architectures.}
HSTU~\cite{Zhai2024ActionsSpeakLouder} established a new paradigm of generative recommender with transformers as the dominant backbone, with subsequent works scaling capacity~\cite{Ye2025FuXiAlpha, Ye2025FuXiBeta, Chen2026VISTA, Yan2026LargeUserModel, Pardoe2026CADET}, improving multi-task optimization~\cite{Yang2025MTMD, Zhang2026SMES, Yuan2024PUB, Cao2025xMTF}, or further productionize the idea~\cite{Han2025MTGR, Hertel2025MFALCON}.
OnePiece~\cite{Dai2025OnePiece} applied unified multi-task training on a shared backbone for cascade ranking at Shopee.
RelayGR~\cite{Wang2026RelayGR} enables KV cache reuse across pipeline stages but assume independently trained models.
Each of these methods achieves partial unification, whether in architecture (HSTU, OnePiece), training (OnePiece), or serving (RelayGR), but none unifies input, model, training and serving end-to-end.

\paragraph{End-to-end generative methods.}
OneRec~\cite{Deng2025OneRec},  OneRanker~\cite{Sun2026OneRanker}, and GPR~\cite{Zhang2025GPR} replace the cascaded pipeline with generative models that produce candidates via autoregressive decoding.
This paradigm has been extended to unified search and recommendation~\cite{Gao2025SynerGen, Chen2025UniSearch, Yan2025IntSR} and deployed across multiple companies~\cite{Han2025MTGR, Sun2026GRank, Zou2026GenRec, Su2026STATIC}.
Beyond infrastructure concerns, recent work~\cite{Yang2024LIGER, ding2026grgeneral} has shown that generative retrieval with semantic ID is not a clear win over dense retrieval due to lossy compression, inefficiency of the generation objective, and cold-start failures, also token relations are memorized in-parameter rather than delegated to external scalable indices where embedding proximity is naturally preserved.


\section{Methodology}\label{sec:method}

\subsubsection{Background:}
\label{sec:method:recgpt-retrieval}

\begin{figure*}[t]
\centering
\includegraphics[width=0.9\textwidth]{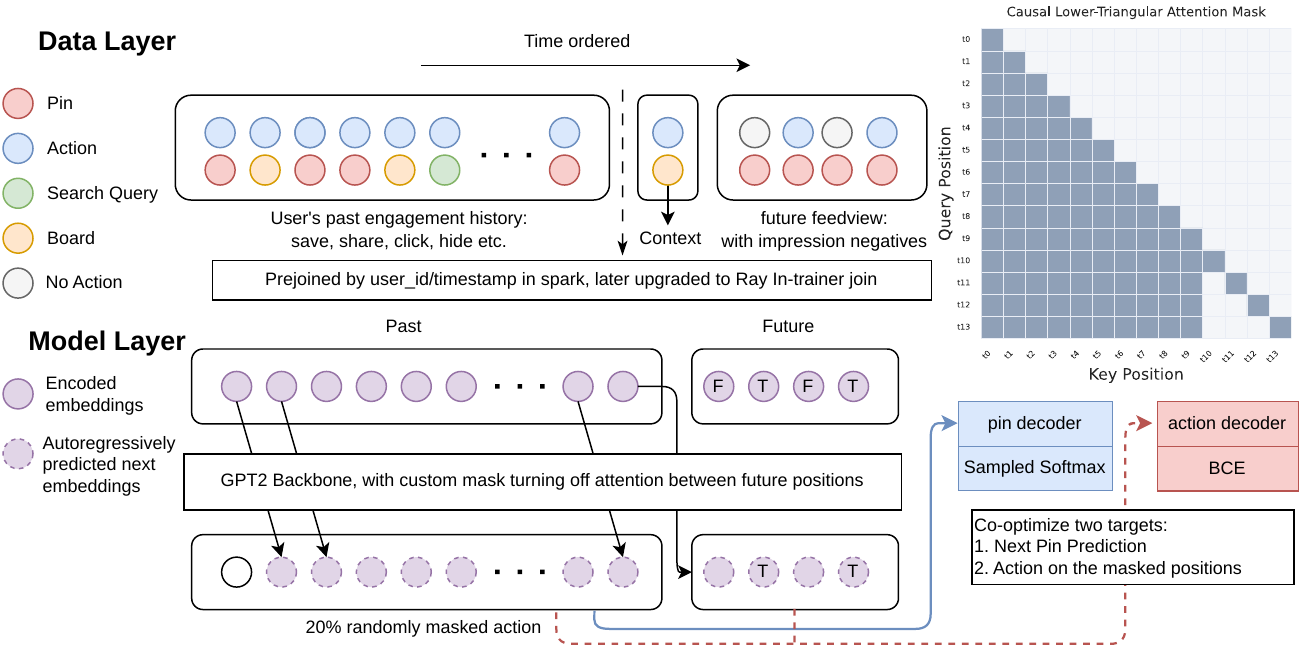}

\caption{End-to-end flow of UniPinRec: a single shared backbone is trained jointly on retrieval and ranking objectives, then co-served as two Triton processes that share the user-history KV cache across stages.}
\label{fig:model_flow}
\end{figure*}

UniPinRec builds on PinRec~\cite{Agarwal2025PinRec}, a generative retrieval model. Each item is represented by pre-trained item embeddings Omnisage\cite{badrinath2025omnisage}, and CLIP based multimodal embeddings\cite{beal2026pinclip}, search queries use pre-trained search embeddings\cite{agarwal2024search}. These features are projected through an MLP into an $L_2$-normalised embedding. Given a user's history of positively interacted items $H(u, t_{\max})$, PinRec feeds the embedding sequence into a causal decoder-only transformer and trains with a sampled softmax loss over in-batch and random negatives:
\begin{equation}
    L_s(\hat{i}_{u,t}, i_{u, t+1}) = -\log \frac
        {\exp\bigl(s(\hat{i}_{u,t}, i_{u, t+1})\bigr)}
        {\exp\bigl(s(\hat{i}_{u,t}, i_{u, t+1})\bigr) + \sum_{i_n \in N} \exp\bigl(s(\hat{i}_{u,t}, i_n)\bigr)},
\end{equation}
where $s(\hat{i}_{u,t}, i_c) = \lambda \cdot \hat{i}_{u,t}^{\top} i_c - \log Q(i_c)$ is a frequency-corrected similarity, $Q(i_c)$ is the estimated sampling probability of candidate $i_c$ (computed via a count-min sketch), and $N$ is the set of negative samples. The model is optimized to minimize $L_s$ averaged over all positions in the history.

\paragraph{Section roadmap.}
The remainder of this section addresses each challenge from the introduction in turn: Section~\ref{sec:method:reformulate} resolves architectural and training divergence via Masked Action Modeling and a joint loss; Section~\ref{sec:method:data} describes the data infrastructure that makes joint training feasible at scale; Section~\ref{sec:method:serving} resolves serving and operational challenges via cross-request KV-cache sharing.

\subsection{Reformulating PinRec for ranking}
\label{sec:method:reformulate}

Extending PinRec to ranking requires three changes:
\begin{enumerate}[nosep,leftmargin=*]
    \item \textbf{Objective:} rather than maximizing next-item likelihood over the full corpus, the model must predict per-candidate action probabilities over a small impression set.
    \item \textbf{Targets:} the supervision shifts from a softmax over item embeddings to binary labels for each action type $c \in \{1,\dots,C\}$ (click, save, hide, etc.), trained with per-head binary cross-entropy.
    \item \textbf{Data:} training examples must include impression negatives (items shown but not engaged with), not just positive engagements (detailed in Section~\ref{sec:method:data}).
\end{enumerate}

\subsubsection{Masked Action Modeling}
\label{sec:method:mam_non_interleave}

The central question is how to formulate action prediction as a sequence learning problem so that the ranking objective can share the retrieval backbone.
The HSTU~\cite{Zhai2024ActionsSpeakLouder} approach interleaves action tokens between items to ensure action predictions are candidate-aware, but this doubles context length and breaks input-format compatibility with retrieval.

We propose \emph{Masked Action Modeling} (MAM): actions are concatenated with each item embedding along the feature dimension and randomly masked during training, preserving causality (the model cannot see an action before predicting it) without inflating the sequence.
A similar mask-and-predict idea appears in~\cite{Huang25Masked} for feed impressions. To our knowledge, we are the first to apply action supervision directly to the user-sequence backbone used for retrieval. This achieves ranking within the same forward pass while simultaneously improving retrieval quality through denser per-position gradients.

\begin{enumerate}
    \item \textbf{Input representation.}
    Rather than interleaving a dedicated action token after every item, we encode the action multi-hot vector $a_i$ as a dense embedding via a linear layer and concatenate it with the item representation along the feature axis before the input projector.
    The projector maps the concatenated representation into the transformer hidden dimension.

    \item \textbf{Masking schedule.}
    For positions in the user's observed history (past), each action is independently masked with probability $p_{\text{mask}}$.
    For candidate (future) positions, actions are always masked ($m_j = 0$ for all $j$), since at inference time the actions on candidates are unknown.
    
    When a position is masked, its action-type input is replaced with a dedicated \texttt{[MASK]} class (an additional one-hot dimension appended to the action vocabulary), so the model can distinguish ``action unknown'' from ``no action taken.''
    
    This guarantees no information leakage. Input masking ($m_i{=}0$) replaces $a_i$ with a \texttt{[MASK]} token, and causal attention restricts position $i$ to $\{0,\ldots,i\}$, so the hidden state satisfies:
    \begin{equation}
    \label{eq:mam-causality}
        z_i = f_\theta\!\bigl(\,\tilde{\Phi}_i,\;\{\tilde{\Phi}_t\}_{t < i}\bigr),
        \quad
        \tilde{\Phi}_i \perp\!\!\!\perp a_i \;\text{ when } m_i = 0,
    \end{equation}
    ensuring $h_{\psi_c}(z_i)$ depends only on $\Phi_i$ and $\{(\Phi_t, m_t \odot a_t)\}_{t<i}$.

\item \textbf{Attention pattern.}
    The transformer processes the concatenated sequence $[\tilde{\Phi}_1, \ldots, \tilde{\Phi}_n, || \tilde{\Phi}'_1, \ldots, \tilde{\Phi}'_k]$ with a modified causal mask (Figure~\ref{fig:model_flow}, upper right) following the M-FALCON pattern in \cite{Zhai2024ActionsSpeakLouder}: past positions retain standard causal attention; each candidate position $n+j$ attends to all past positions $1, \ldots, n$ but is blocked from attending to other candidates.
    All candidate positions share fixed position IDs, feedview types and timestamps to ensure unbiased scoring.
    Because candidates cannot attend to each other, the mask reduces attention cost from $\mathcal{O}((n{+}k)^2)$ to $\mathcal{O}(n^2 + nk)$; we materialise this sparsity via \emph{flex attention}~\cite{he2024flexattention}, which compiles the block-sparse pattern into fused CUDA kernels.
    This pattern also enables KV-cache sharing across stages (Section~\ref{sec:method:kv_cache}): retrieval caches the $\mathcal{O}(n^2)$ history computation, and ranking reuses it at a marginal cost of $\mathcal{O}(nk)$.

    \item \textbf{Training objective.}
    The past portion of the sequence $\tilde{\Phi}_1, \ldots, \tilde{\Phi}_n$ retains the same length and positional structure as the retrieval model's input, and is in fact used for retrieval: the same next-item prediction objective is applied over these positions. The $k$ candidate items $\tilde{\Phi}'_1, \ldots, \tilde{\Phi}'_k$ are appended after the past, and the full sequence is processed in a single forward pass. This enables full parameter sharing with the retrieval model. The model is trained with two jointly optimised losses:
    \begin{enumerate}
        \item \textbf{Next-item prediction loss} (retrieval objective): identical to the sampled-softmax loss in Section~\ref{sec:method:recgpt-retrieval}, applied at every unmasked past position:
        \begin{equation}
            \mathcal{L}_{\text{item}} = \frac{1}{|H|}\sum_{t \in H} L_s\!\bigl(\hat{i}_{u,t},\; i_{u,t+1}\bigr).
        \end{equation}
        \item \textbf{Action prediction loss} (ranking objective): at every position where the action was masked, the model must predict each action type independently. For each action type $c \in \{1,\dots,C\}$, a dedicated MLP head $h_{\psi_c}$ produces a scalar logit, and a per-head weight $w_c$ controls its contribution. Separate terms are applied to past (masked) and future (always masked) positions:
        \begin{equation}
        \begin{split}
            \mathcal{L}_{\text{action}} = \sum_{c=1}^{C} w_c \Bigl[\;
            &\frac{1}{|\mathcal{M}_{\text{past}}|}
              \!\sum_{i \in \mathcal{M}_{\text{past}}}
              \ell_{\text{BCE}}\!\bigl(h_{\psi_c}(z_i),\, a_i^{(c)}\bigr) \\
            &+ \frac{1}{k}\sum_{j=1}^{k}
              \ell_{\text{BCE}}\!\bigl(h_{\psi_c}(z'_j),\, a'^{(c)}_j\bigr)
            \Bigr],
        \end{split}
        \end{equation}
        where $C$ is the number of action types, $z_i$ is the transformer hidden state at position $i$, $h_{\psi_c}$ is the MLP head for action type $c$, $a_i^{(c)}$ is the binary label for action $c$ at position $i$, $\mathcal{M}_{\text{past}} = \{i : m_i = 0\}$ is the set of masked past positions, and $w_c$ are hyperparameter-tuned scalar weights that balance the contribution of each action type (e.g., save, click, hide) and against next-item prediction loss.
    \end{enumerate}

    The total loss is:
    \begin{equation}
        \mathcal{L} = \mathcal{L}_{\text{item}} + \mathcal{L}_{\text{action}}.
    \end{equation}

\begin{figure*}[!t]
\centering
\includegraphics[width=0.95\textwidth]{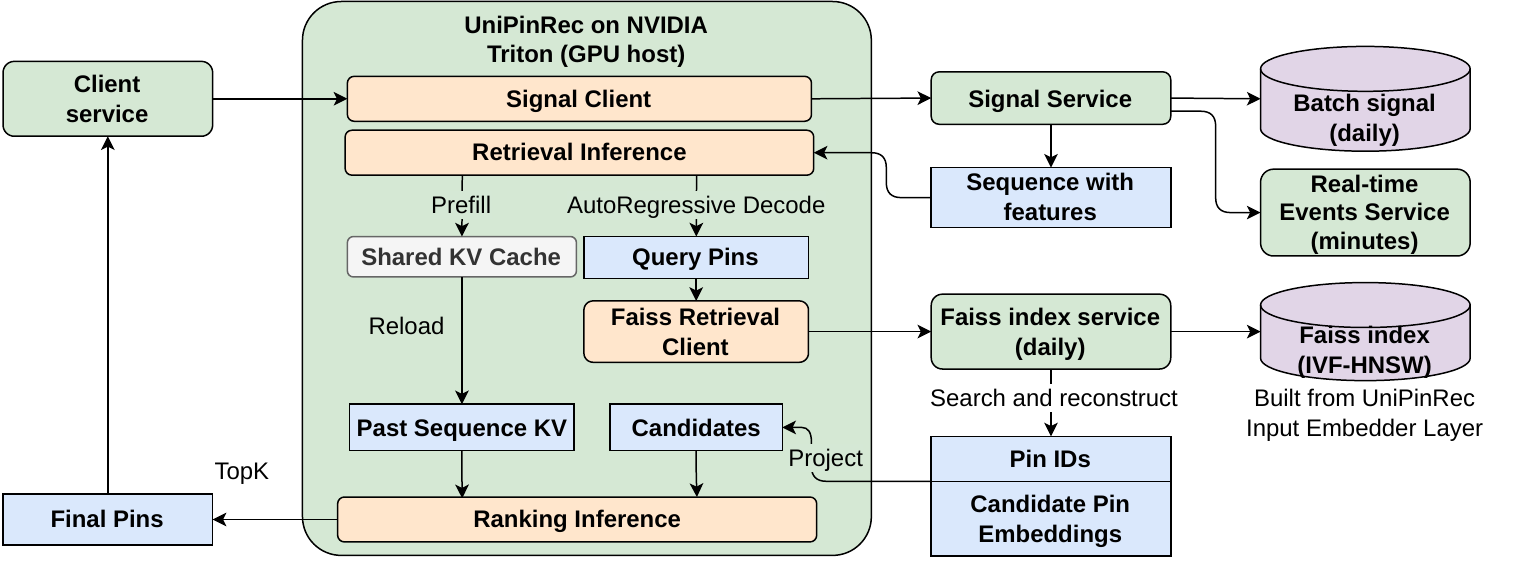}
\caption{Co-serving topology for UniPinRec. Retrieval and ranking are independent Triton ensemble nodes; an ANN lookup against a CPU-hosted Faiss index sits between them. ANN returns both candidate Pin IDs and their reconstructed item embeddings, sparing the ranking stage a separate embedding fetch. The shared KV pool is reused by ranking.}
\label{fig:serving_flow}
\end{figure*}

    \item \textbf{Key advantages over interleaving.}
    \begin{enumerate}[nosep,leftmargin=*]
        \item No context-length inflation: the sequence length stays identical to the retrieval model.
        \item Full weight sharing: the ranking model can be initialised directly from a retrieval checkpoint.
        \item Denoising regularisation: random masking encourages the model to rely on item content when actions are unavailable, improving generalisation at inference time.
    \end{enumerate}

\end{enumerate}

\subsection{Data infrastructure}
\label{sec:method:data}

Joint training introduces a data requirement absent from single-task models: each training example must contain both the user's engagement history (for the retrieval objective) and the full impression slate with negative outcomes (for the action prediction objective).
Neither existing retrieval nor ranking datasets satisfy both constraints simultaneously.
Constructing this unified format at scale
is a prerequisite for full-stack unification.
We describe how we construct and serve these examples below.

\subsubsection{Feedview-centric dataset construction}
\label{sec:method:feedview}

A key data design choice distinguishes our work from prior sequential recommenders.
Retrieval models typically train on action-only sequences (chronological engagements)~\cite{Zhai2024ActionsSpeakLouder, Agarwal2025PinRec, Deng2025OneRec}, while ranking models train on impression logs (the full slate shown to a user, with both engaged and no-action outcomes)~\cite{Hertel2025MFALCON, Dai2025OnePiece}.
Prior work on unified models has not discussed how to construct training data that satisfies both objectives; joint training requires both: the retrieval objective needs positive action history to predict the next item, while the ranking objective needs the full impression slate to learn. 

We bridge this by constructing each training example with a \emph{past} action sequence paired with a \emph{future} feedview (the full impression slate from a subsequent request, with engagement labels). 
Concretely, each position in the sequence is a (Pin, surface, action, timestamp) tuple, where action spans the engagement vocabulary the model predicts (impression-only, click, save, share, hide, etc.).

Because engagement events are naturally sparse relative to impressions, the training pipeline subsamples non-engaged feedviews (retaining $10\%$) to amplify rare engagement classes. The evaluation pipeline applies no downsampling and uses a randomized replay dataset, preserving the natural unbiased distribution. This also benefits retrieval by reflecting the real traffic distribution.

\subsubsection{Ray in-trainer join}
\label{sec:method:ray-in-trainer-join}

We eliminate the user action history fanout introduced in offline join by performing an \emph{in-trainer bucket join} using Ray~\cite{moritz2018ray} distributed dataloading layer. User action sequences and feedview records are maintained as separate Iceberg tables, each hash-bucketed by user ID, and joined in memory at training time. 
This avoids data duplication entirely and makes context length, sampling ratios, and filtering tweakable parameters at training time instead of data-generation time.

\subsection{Serving infrastructure}
\label{sec:method:serving}

Figure~\ref{fig:serving_flow} shows the end-to-end co-serving topology described in this section.

A unified model only delivers efficiency benefits if it can replace two separate models without doubling serving costs.
Two properties of our model design/training protocol make this tractable.
First, the non-interleaved architecture and unified co-training (Section~\ref{sec:method:mam_non_interleave}) guarantee the same model weights and produce candidate independent user representations that can be cached once and reused across stages.
Second, both the retrieval and ranking stages share the same item embedding space, so there's no need to maintain separate candidate representations.

\subsubsection{Background:}
\label{sec:method:pinrec_serving}

We build on the production PinRec~\cite{Agarwal2025PinRec} serving stack, which uses a Triton ensemble to assemble user history, run autoregressive retrieval, and perform ANN lookup through a CPU-hosted Faiss index.

\subsubsection{Overall serving design}
\label{sec:method:triton}

UniPinRec retains most of this stack and attaches a new ranking node downstream of Faiss.
The ranking node consumes the ANN's candidate set along with a KV-cache slot identifier (detailed in Section~\ref{sec:method:kv_cache} below).
Rather than reprocessing the full user history, it runs an incremental forward pass: only the candidate tokens are processed, attending back to the KV cache history to produce per-candidate action probabilities.

The ANN index already stores each Pin's embedded vector (the output of the embedder MLP, computed at index-build time).
We use \texttt{Faiss.search\_and\_reconstruct}, which returns both the candidate Pin IDs and their stored embeddings in a single call.
These pre-embedded vectors serve directly as candidate tokens for the ranking pass.
This eliminates additional embedder forward pass and feature fetches on the candidate side, and guarantees that the retrieval and ranking stages score against identical representations.

\subsubsection{KV-cache sharing across stages}
\label{sec:method:kv_cache}

The dominant cost in our transformer is encoding the user's action history ($n$ tokens, $\mathcal{O}(n^2)$ self-attention); scoring $k$ candidates against that history is only $\mathcal{O}(nk)$.
We avoid the $\mathcal{O}(n^2)$ cost entirely during ranking by extending PinRec's per-request KV cache (Section~\ref{sec:method:pinrec_serving}) across stages and across processes: ranking attends directly to the cached history KV without re-encoding it.
We further prevent candidates from attending to each other, following the M-FALCON~\cite{Hertel2025MFALCON} identity-mask design, so each candidate's score is independent of the batch composition.
The net ranking cost is $\mathcal{O}(nk)$.

Retrieval and ranking run as separate OS processes with separate CUDA contexts, as is typical in microservice-based serving stacks.
Naive serialization of KV tensors across processes incurs GPU-host-GPU roundtrips that dominate latency.
We instead pre-allocate a GPU memory pool that both processes map into their address spaces, so the ranking process reads the KV state written by retrieval without CPU to GPU data transfer.

Each retrieval replica therefore allocates a pre-sized GPU pool of shape $[L, S, H, n, D]$ (layers, slots, KV heads, past sequence length, head dim), exports CUDA IPC handles for its storage to a per-instance file at boot time, and the ranking process opens those handles and maps the same physical GPU allocation into its own address space.

Slots are reused round-robin, 
each ranking consumer copies each request's past KV from its slot in the mapped pool into its own decode cache during the forward pass.

\subsubsection{FP8 quantization}
\label{sec:method:fp8}
To further improve serving efficiency, we use mixed-precision FP8 training and inference. We use fused modules that combine layer normalization, linear projection, and activation into single kernels from the Transformer Engine library, avoiding intermediate full-precision materialization. Training uses a hybrid FP8 format with E4M3 during the forward pass and E5M2 for backward computation so that forward activations and backward gradients retain an appropriate balance between precision and dynamic range. Quantization comes at a cost of a 0.5\% drop in offline metrics, which we accepted as a reasonable tradeoff for the efficiency gains.

\section{Results}\label{sec:experiments}

We structure our results around three research questions:
\begin{itemize}
    \item \textbf{RQ1}: Can we train a unified model that is capable of both retrieval and ranking tasks?
    \item \textbf{RQ2}: Can we co-serve a retrieval and ranking model to achieve minimal compute redundancy?
    \item \textbf{RQ3}: Can we make this unified model an easy extension to existing production level recommendation system?
\end{itemize}

\subsection{Train a Unified Retrieval and Ranking (RQ1)}
We demonstrate the generative recommendation paradigm can be extended to ranking model. This requires (1) architectural compatibility across stages to enable parameter sharing, and (2) joint training with both retrieval and ranking objectives to allow signal sharing between tasks.

\paragraph{Baselines and training strategies.}
We compare UniPinRec against multiple baselines and training approaches:

\textbf{TransAct V2 + DCNv2} \cite{xia2025transactv2}: The production ranking model contemporarily trained, a transformer based model with long user sequence and other feature crossing modules. 

\textbf{HSTU}~\cite{Zhai2024ActionsSpeakLouder}: State-of-the-art unified retrieval and ranking model with action and item interleaving \cite{twobird2025}. 
We evaluate HSTU on both retrieval and ranking tasks with unified interleaved paradigm with matching effective sequence length.

\textbf{PinRec}~\cite{Agarwal2025PinRec}: The production retrieval model, a 12-layer transformer trained with next-item prediction (sampled softmax). Serves only retrieval; no ranking capability. 

\textbf{PinRec finetuned}: Sequential training approach where we take the snapshot above (PinRec), then finetune for ranking by adding action prediction objectives. Critically, the item embedder must remain frozen during finetuning to maintain compatibility with the pre-trained retrieval model's Faiss index for deployment.

\textbf{UniPinRec w/o item loss}: A ranking-only variant trained exclusively with action prediction loss, using the same MAM architecture as the unified model. This ablation isolates the effect of joint training by removing the retrieval objective.

\textbf{UniPinRec (ours)}: The unified model with Masked Action Modeling, using the same non-interleaved sequence format for both retrieval and ranking. Trained jointly with both next-item prediction (retrieval) and action prediction (ranking) losses. The item embedder is trainable and benefits from both objectives.

\paragraph{Results.}
Table~\ref{tab:unified} shows that UniPinRec achieves competitive performance on both retrieval (Recall@10, interacted targets in the future feedview that are correctly retrieved in the top 10 from any autoregressively generated query embeddings) and ranking ("save" Hit@3, ranked by save head score) compared to specialized baselines. We use "Board More Ideas" feed data to train and evaluate. More online results will be discussed in Section~\ref{sec:online_results:bmi_online}

\begin{table}[H]
\caption{Unified joint training vs.\ detached pre-train / fine-tune. Both approaches use the same 12-layer MAM backbone. Two columns are on different evaluation datasets. Hit@3 measures ranking performance and Recall@10 measures retrieval performance. Best result in \textbf{bold}.}
\label{tab:unified}
\begin{tabular*}{\columnwidth}{@{\extracolsep{\fill}}p{0.38\columnwidth}cc@{}}
\toprule
\textbf{Strategy} & \textbf{Hit@3} $\uparrow$ & \textbf{Recall@10} $\uparrow$ \\
\midrule
TransActV2 + DCNv2 & 0.088008 & N/A \\
HSTU & 0.097326 & 0.76161 \\
PinRec & N/A & 0.77486 \\
PinRec finetuned & 0.095869 & N/A \\
UniPinRec w/o item loss & 0.097345 & N/A \\
UniPinRec (ours)        & \textbf{0.10096} & \textbf{0.77659} \\
\bottomrule
\end{tabular*}
\end{table}

On retrieval, UniPinRec matches PinRec's Recall@10, with \textbf{+0.2\%} marginal improvement, demonstrating that joint training with ranking objectives does not degrade retrieval quality. This is critical for deployment: the unified model can serve as a drop-in replacement for the production retrieval system.

On ranking, UniPinRec surpasses TransActV2+DCNv2, the contemporary production ranking model by \textbf{14.7\%} with dedicated feature crossing modules, despite handling both retrieval and ranking in a single model. UniPinRec also outperforms HSTU at the same context length budget. Importantly, UniPinRec outperforms the ranking-only variant (UniPinRec w/o item loss), showing that the joint retrieval objective further improves ranking quality. The PinRec fine-tuned approach underperforms joint training despite starting from a strong retrieval model. The frozen embedder constraint prevents the model from adapting item representations for ranking, while joint training allows the embedder to learn from both tasks simultaneously. This gap demonstrates that architectural compatibility alone is insufficient, joint optimization is necessary to fully leverage unified training.

\subsection{Co-serve a Retrieval and Ranking Model (RQ2)}
A unified backbone can only claim efficiency win if it does not serve as two disjoint models.
The dominant compute in a ranking forward pass is the self-attention over the user's history; when retrieval and ranking are served separately, this cost is paid twice per request.
Our unified design encodes the history once during retrieval and reuses the resulting keys and values as the KV cache for the subsequent ranking pass, turning ranking into a \emph{decode} step that attends from candidates into a frozen history rather than a full \emph{prefill}.

We measure the impact of this reuse in Table~\ref{tab:serving_latency}, on NVIDIA L40S ($B{=}8$, $n{=}992$, $k{=}656$, $L{=}12$), varying three orthogonal levers in addition to the path: the activation dtype (bf16 vs.\ fp8 with Transformer~Engine fused matmuls), the attention implementation (SDPA vs.\ auto-tuned flex attention), and the runtime (eager Python dispatch, \texttt{torch.compile}, or CUDA graph capture)
:
\begin{itemize}
    \item KV-cache reuse (prefill$\to$decode) roughly $\sim$$2.4\times$ .
    \item fp8 with Transformer~Engine matmuls and CUDA graph capture gives a $\sim$$1.25\times$ over bf16+compile.
    \item Flex attention contributes to a steady $\sim$$1.3\times$ on top of any dtype.
\end{itemize}
The three levers are largely orthogonal, so stacking them yields the maximum end-to-end $\mathbf{3.92\times}$ in the bottom-right cell of Table~\ref{tab:serving_latency}.

\begin{table}[h!]
\caption{Eval-time forward latency for a single ranking pass
Speedups are relative to the bf16 / SDPA / compile prefill row. ``Runtime'' controls how the forward is dispatched: eager, torch.compile or cuda graph capture. We use graph for fp8 decode because torch.compile fusion regresses Transformer~Engine kernels. Best result in \textbf{bold}, *used in online experiments}\label{tab:serving_latency}
\begin{tabular*}{\columnwidth}{@{\extracolsep{\fill}}lllrr@{}}
\toprule
\textbf{Dtype} & \textbf{Attn} & \textbf{Runtime} & \textbf{Lat.\,(ms)} $\downarrow$ & \textbf{Speedup} $\uparrow$ \\
\midrule
\multicolumn{5}{l}{\emph{Prefill (history recomputed)}} \\
bf16 & SDPA & compile & 25.72 & $1.00\times$ \\
bf16 & flex & compile & 19.70 & $1.31\times$ \\
fp8  & SDPA & eager & 18.92 & $1.36\times$ \\
fp8  & SDPA & compile & 18.91 & $1.36\times$ \\
\midrule
\multicolumn{5}{l}{\emph{Decode (KV-cache reuse, GPU memcpy overhead included)}} \\
bf16 & SDPA & compile & 10.42 & $2.47\times$ \\
bf16 & flex & compile* & \phantom{0}8.57 & $3.00\times$ \\
fp8  & SDPA & graph   & \phantom{0}8.77 & $2.93\times$ \\
fp8  & flex & graph   & \textbf{\phantom{0}6.56} & $\mathbf{3.92\times}$ \\
\bottomrule
\end{tabular*}
\end{table}

\begin{table}[h!]
\caption{Online serving wins at controlled GPU activity/budget. Baseline uses (bf16, SDPA, torch.compile) for both retrieval and ranking, and all numbers are percentage changes relative to this joint baseline. *used in online experiments in Section~\ref{sec:online}}

\label{tab:serving_qps}
\begin{tabular*}{\columnwidth}{@{\extracolsep{\fill}}lcc@{}}
\toprule
\textbf{Variant} & \textbf{E2E latency} $\downarrow$ & \textbf{QPS lift} $\uparrow$ \\
\midrule
baseline                      & --- & --- \\
+ bf16 decode flex compile* & \textbf{$-11.1\%$} & \textbf{$+63.6\%$} \\
+ fp8 decode flex graph & \textbf{$+6.7\%$} & \textbf{$+109.1\%$} \\
\bottomrule
\end{tabular*}
\end{table}

We further benchmark the best groups of bf16 and fp8 in online conditions Table~\ref{tab:serving_qps}.
One major trade-off we observed with fp8 kernels is the necessity to wait and accumulate requests to form larger batch size to meet the shape constraints of Transformer Engine where the flattened leading dimension $[B * S, D]$ must be a multiple of 8, this mostly constrains the autoregressive steps of the retrieval forward pass where $S = 1$. Resulting in increased QPS but at larger latency. For the online experiments reported, we used bf16, but plan to proceed with fp8 as next steps.   

\subsection{Ablations}

\subsubsection{Effect of Masking Ratio in MAM}
\label{sec:ablation_masking}
A core design choice in Masked Action Modeling (MAM) is the masking probability $p_{\text{mask}}$, which controls how often historical actions in the user sequence are randomly masked during training. We ablate this choice by training variants with $p_{\text{mask}} \in \{0.0, 0.1, 0.2, 0.3\}$ on identical data and hyperparameters, holding all other architectural decisions constant.

\paragraph{Results.}
Table~\ref{tab:masking_ratio} shows that \emph{any} masking improves over no masking ($p_{\text{mask}} = 0$). The zero-masking baseline achieves the lowest Hit@3, confirming that additional action prediction signal and regularization both contribute to better ranking quality.
with $p_{\text{mask}} = 0.2$ yielding the best overall performance. Higher masking ($p_{\text{mask}} = 0.3$) shows diminishing returns, likely because overly aggressive masking removes too much signal from the user history.

\begin{table}[h!]
\caption{Effect of action masking ratio $p_{\text{mask}}$ in MAM. Best result in \textbf{bold}.}
\label{tab:masking_ratio}
\begin{tabular*}{\columnwidth}{@{\extracolsep{\fill}}lc@{}}
\toprule
\textbf{$p_{\text{mask}}$} & \textbf{Hit@3} $\uparrow$ \\
\midrule
0.0 (no masking)  & 0.09923  \\
0.1               & 0.10078  \\
0.2               & \textbf{0.10096} \\
0.3               & 0.10075  \\
\bottomrule
\end{tabular*}
\end{table}

\subsubsection{Effect of Model Scaling}
\label{sec:ablation_scaling}
\begin{figure}[h!]
  \centering
  \begin{tikzpicture}
  \begin{axis}[
    width=\columnwidth,
    height=4.5cm,
    xlabel={Relative FLOPs ($1{\times}$ = smallest model)},
    ylabel={Hit@3},
    xmin=0, xmax=15.5,
    ymin=0.095, ymax=0.1045,
    ymajorgrids=true,
    grid style=dashed,
    legend pos=south east,
    legend style={font=\small},
    tick label style={font=\small},
    label style={font=\small},
    scaled y ticks=false,
    yticklabel style={/pgf/number format/fixed, /pgf/number format/precision=4},
  ]
  \addplot[color=sbblue, mark=square*, thick] coordinates {
    (1.0,  0.09584)
    (2.0,  0.09855)
    (4.0,  0.10019)
    (6.0,  0.10096)
    (12.0, 0.10147)
  };
  \addlegendentry{Depth (L=2,4,8,12,24; S=1024)}

  \addplot[color=sborange, mark=*, thick] coordinates {
    (1.3,  0.09647)
    (2.7,  0.09829)
    (6.0,  0.10096)
    (14.2, 0.10353)
  };
  \addlegendentry{Seq length (S=256,512,1K,2K; L=12)}

  \end{axis}
  \end{tikzpicture}
  \caption{Hit@3 vs.\ relative FLOPs for depth and sequence length scaling. FLOPs are normalized to the smallest model ($L{=}2$, $S{=}1024$). Both series use MAM with $p_{\text{mask}}=0.2$ .}
  \label{fig:scaling}
\end{figure}
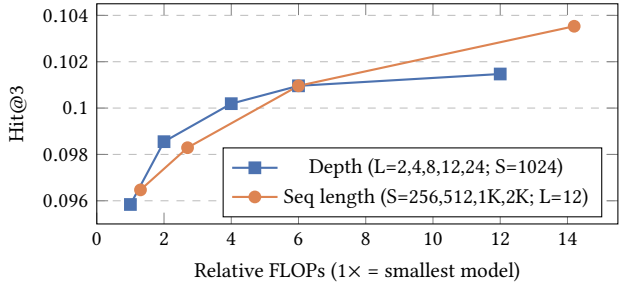

We study how model capacity affects ranking performance across two dimensions: transformer depth and sequence length. 
Understanding these tradeoffs is critical for deployment, as both dimensions impact model quality, training cost, and serving latency.

\paragraph{Results.}
We train models with varying depth (2, 4, 8, 12, 24 layers) holding sequence length fixed at 1024 tokens, and varying sequence length (256, 512, 1024, 2048 tokens) with a fixed 12-layer model. Both dimensions use MAM with $p_{\text{mask}} = 0.2$. Figure~\ref{fig:scaling} shows that performance improves consistently across both scaling axes, demonstrating that the unified model has not saturated in capacity. However, the cost implications differ significantly: depth scales linearly with compute, while sequence length incurs quadratic costs for attention. This makes sequence length a more severe bottleneck for latency-sensitive applications despite both dimensions showing continued quality improvements. 
The strong gains from longer sequences suggest that sequence compression techniques are a promising direction for capturing the quality benefits of longer context at reduced cost.

\section{Online A/B Experiments (RQ3)}\label{sec:online}
We evaluate UniPinRec in production through controlled A/B experiments, comparing the unified model against PinRec~\cite{Agarwal2025PinRec}, the production retrieval baseline. This validates whether full-stack unification can deliver engagement improvements in real-world traffic while maintaining practical serving efficiency, and addresses the third RQ, whether we could make this highly modularized extension to existing production recommendation systems.

\subsection{Experimental Setup}

\paragraph{Deployment configurations.}
We compare two setups for candidate generator, both returning K(<3000) candidates to downstream stages:

\textbf{Baseline (PinRec):} The production retrieval model generates K candidates via ANN search, which are passed directly to downstream ranking stages.

\textbf{UniPinRec:} The unified model first overfetches candidates via ANN search (L0), then immediately ranks them using the ranking head (L1) with action prediction, and returns the top K ranked candidates(based on head utility functions) to downstream stages. 

It is worth noting that for the online experiments, we replace retrieval stage with a unified L0 + L1 scoring server and still send the refined candidates to a downstream production ranker based on TransAct V2~\cite{xia2025transactv2} to rigidly validate the metric improvements. This also means the offline lifts will be attenuated and diluted.

We run A/B experiments on two production surfaces: Board More Ideas (L0 overfetch ratio \textasciitilde2x) and Notification (L0 overfetch ratio \textasciitilde3x). Both surfaces serve millions of users daily.

\subsection{Board More Ideas}
\label{sec:online_results:bmi_online}
Board More Ideas (BMI) recommends Pins anchored to a user's board. 
To enable this, we extend UniPinRec to support ``Board'' as a new sequence modality beyond Pins and search queries. 

\paragraph{Results.}
Table~\ref{tab:online_bmi} shows that UniPinRec provides strong engagement lifts over the PinRec baseline with \textbf{+0.95\%} surface save lift, demonstrating that the unified retrieval and ranking approach significantly improves candidate quality over retrieval alone, also driving \textbf{+0.08\%} site-wide lift, confirming that better BMI recommendations positively impact overall platform engagement. 

\begin{table}[h!]
\caption{Online A/B results on Board More Ideas. Lifts show UniPinRec L0+L1 vs PinRec baseline.}
\label{tab:online_bmi}
\begin{tabular*}{\columnwidth}{@{\extracolsep{\fill}}p{0.6\columnwidth}r@{}}
\toprule
\textbf{Metric} & \textbf{UniPinRec} \\
\midrule
BMI surface saves       & \textbf{+0.95\%} \\
Site-wide saves & \textbf{+0.08\%} \\
\bottomrule
\end{tabular*}
\end{table}

\subsection{Notifications}
Pinterest uses email and push notifications to deliver relevant content to users. These notifications are an important lever for user reactivation and long-term retention. 

\paragraph{Results.}
Table~\ref{tab:online_notif} shows that replacing the PinRec retrieval-only baseline with the UniPinRec ranker delivers consistent gains across notification engagement metrics, with the largest lift on notification-surface saves. The dormant-user push-open lift of \textbf{+1.72\%} is roughly twice the all-user lift, suggesting that more personalized notification recommendations are particularly effective at bringing lapsed users back and noticeably improving WAU by \textbf{0.09\%}.

\begin{table}[H]
\caption{Online A/B results on notifications. Lifts show UniPinRec vs PinRec baseline, measured on a unique-user basis.}
\label{tab:online_notif}
\begin{tabular*}{\columnwidth}{@{\extracolsep{\fill}}p{0.6\columnwidth}r@{}}
\toprule
\textbf{Metric} & \textbf{UniPinRec} \\
\midrule
Push opens                  & \textbf{+0.91\%} \\
Push opens (dormant users)  & \textbf{+1.72\%} \\
Notification surface saves & \textbf{+3.84\%} \\
Email clicks                & \textbf{+0.30\%} \\
Weekly active users        & \textbf{+0.09\%} \\

\bottomrule
\end{tabular*}
\end{table}

\section{Conclusion}\label{sec:conclusion}
We presented UniPinRec, the first full-stack unification of retrieval and ranking deployed in a production recommendation system.
Empirically, the unified model matches production retrieval recall, holds ranking quality exceeding dedicated ranker, and cuts e2e latency while lifting QPS,  surfacing candidates the previous funnel missed and translating into a net engagement win online.

UniPinRec currently unifies retrieval with upper-funnel ranking, where backbone redundancy is most severe.
It does not yet replace the L2 ranker, which has additional infrastructure dependencies in score calibration, training data logging, and operational tooling beyond the immediate scope of this work. We take an incremental rollout approach and plan to extend the unification effort in L2 stage and horizontally to additional surfaces (Search, Ads) as next steps.

Architecturally, UniPinRec mirrors retrieval-augmented generation: the backbone formulates a query, an ANN index serves as non-parametric memory, and the ranker generates action probabilities through the same cached context.
This parallel suggests borrowing from the RAG literature, such as iterative retrieval and learned retrieval-depth control, as future directions.

Prior unified models have demonstrated that retrieval and ranking can share an architecture.
UniPinRec shows that the harder problem, unifying input format, training, and serving infrastructure, is also solvable in production today.


\begin{acks}
We thank collaborators such as Kyle Soares, Pihui Wei, Anirudhan Badrinath, Edoardo Botta, Devin Kreuzer, Josie Zeng, Chi Zhang, Chen Yang, Po-Wei Wang, David Woo, Lei Wang, Lin Zhu, Kritarth Anand, Bowen Deng, Matt Chun, Zisis Petrou, Dimitra Tsiaousi, Dylan Wang for their helpful support, valuable feedbacks and insightful discussions throughout the development of this work.
\end{acks}

\bibliographystyle{ACM-Reference-Format}
\bibliography{submission/base}

\end{document}